\documentclass[reprint,longbibliography,aps,prl]{revtex4-1}
\usepackage[utf8]{inputenc}
\usepackage{amsmath,amssymb,amsthm,mathtools}
\usepackage{graphicx}
\usepackage{dcolumn}
\usepackage{bm}
\usepackage{hyperref}

\def\be{\begin{equation}}
\def\ee{\end{equation}}
\def\ba#1\ea{\begin{align}#1\end{align}}
\def\bg#1\eg{\begin{gather}#1\end{gather}}
\def\bm#1\em{\begin{multline}#1\end{multline}}
\def\bmd#1\emd{\begin{multlined}#1\end{multlined}}

\def\a{\alpha}
\def\b{\beta}

\def\d{\delta}

\def\e{\epsilon}

\def\g{\gamma}

\def\m{\mu}
\def\n{\nu}

\def\r{\rho}

\def\S{\Sigma}
\def\t{\tau}

\def\W{\Omega}

\def\la{\label}

\def\re{\ref}
\def\er{\eqref}
\def\se{\section}

\def\fr{\frac}

\def\pa{\partial}

\def\wtd{\widetilde}

\def\ph{\phantom}
\def\eq{\equiv}

\def\cd{\cdots}
\def\ap{\approx}
\def\nn{\nonumber}
\def\qu{\quad}

\def\lt{\left}
\def\rt{\right}
\def\({\left(}
\def\){\right)}
\def\[{\left[}
\def\]{\right]}
\def\<{\langle}
\def\>{\rangle}

\def\tr{\operatorname{tr}}

\def\bH{{\mathbb H}}

\def\cC{{\mathcal C}}

\def\cK{{\mathcal K}}
\def\cO{{\mathcal O}}

\def\cR{{\mathcal R}}

\def\Area{\operatorname{Area}}

\def\bulk{\text{bulk}}

\def\univ{\text{univ}}

\newcommand{\T}[3]{{#1^{#2}_{\ph{#2}#3}}}
\newcommand{\Tu}[3]{{#1_{#2}^{\ph{#2}#3}}}
\newcommand{\Tud}[4]{{#1^{\ph{#2}#3}_{#2\ph{#3}#4}}}

\begin{document}

\title{Shape Dependence of Holographic R\'enyi Entropy in Conformal Field Theories}
\author{Xi Dong}
\email{xidong@ias.edu}
\affiliation{School of Natural Sciences, Institute for Advanced Study, Princeton, New Jersey 08540, USA}
\date{\today}

\begin{abstract}
We develop a framework for studying the well-known universal term in the R\'enyi entropy for an arbitrary entangling region in four-dimensional conformal field theories that are holographically dual to gravitational theories.  The shape dependence of the R\'enyi entropy $S_n$ is described by two coefficients: $f_b(n)$ for traceless extrinsic curvature deformations and $f_c(n)$ for Weyl tensor deformations.  We provide the first calculation of the coefficient $f_b(n)$ in interacting theories by relating it to the stress tensor one-point function in a deformed hyperboloid background.  The latter is then determined by a straightforward holographic calculation.  Our results show that a previous conjecture $f_b(n) = f_c(n)$, motivated by surprising evidence from a variety of free field theories and studies of conical defects, fails holographically.
\end{abstract}

\maketitle

\se{Introduction}

Quantum entanglement has been playing an increasingly dominant role in understanding complex systems in a diverse set of areas including condensed matter physics \cite{Levin:2006zz, Kitaev:2005dm, Li:2008kda}, quantum information \cite{Bennett:2000quantum}, and quantum gravity \cite{Jacobson:1995ab, Ryu:2006bv, Hubeny:2007xt, VanRaamsdonk:2010pw, Bianchi:2012ev, Maldacena:2013xja, Dong:2013qoa, Faulkner:2013ica, Almheiri:2014lwa, Dong:2016eik}.  One measure of entanglement is the von Neumann entropy for the density matrix of a subsystem, also known as the entanglement entropy.

A different set of measures of entanglement is provided by the R\'enyi entropies $S_n$ labeled by an index $n$, a one-parameter generalization of the von Neumann entropy \cite{Renyi:1961}.  However, they are much easier to experimentally measure \cite{Abanin:2012, Daley:2012, Islam:2015} and numerically study \cite{Hastings:2010, Kallin:2011, Kallin:2014oka} than the von Neumann entropy.  They also contain much richer physical information about the entanglement structure of a quantum state, and knowing R\'enyi entropies for all $n$ allows one to reconstruct the whole entanglement spectrum, i.e.\ the set of eigenvalues of the density matrix.  R\'enyi entropies have been extensively studied in various contexts including spin chains \cite{Franchini:2007eu}, tensor networks \cite{Hayden:2016cfa}, free field theories \cite{Klebanov:2011uf}, conformal field theories (CFTs) \cite{Holzhey:1994we, Calabrese:2009qy}, and gauge-gravity duality \cite{Headrick:2010zt, Hung:2011nu}.  Furthermore, R\'enyi entropy at index $n=1/2$ gives the entanglement negativity which is a measure of the distillable entanglement contained in a quantum state \cite{Vidal:2001}.

In any $d$-dimensional CFT on a generally curved background, the R\'enyi entropy for a spatial region $A$ is ultraviolet (UV) divergent.  Organized by the degree of divergence, the R\'enyi entropy may be written as
\be\la{snuv}
S_n = \g_n^{(0)} \fr{\Area(\S)}{\e^{d-2}} +\cd +S_n^{\univ} +\cd \,,
\ee
where $\S\eq\pa A$ is the entangling surface and $\e$ is a short distance cutoff.  The first set of dots in \er{snuv} denotes terms with subleading power-law divergences.  The term $S_n^{\univ}$ is universal in the sense that it does not depend on the detail of the UV cutoff, whereas coefficients such as $\g_n^{(0)}$ are scheme dependent and nonuniversal. 

In odd spacetime dimensions, the universal term is independent of $\e$ but depends nonlocally on the (intrinsic and extrinsic) shape of the entangling surface.  In even dimensions, however, $S_n^{\univ}$ is proportional to $\ln\e$ and the universal coefficient is a linear combination of conformal invariants built from integrals of local geometric quantities over the entangling surface.

In two dimensions, the universal term is completely determined by the central charge \cite{Holzhey:1994we, Lunin:2000yv, Calabrese:2004eu, Calabrese:2009qy}:
\be
S_n^{\univ} = -\fr{c}{12} \(1+ \fr{1}{n}\) \Area(\S) \ln\e \,.
\ee
In this case the most general region is a union of $m$ intervals, and the area of $\S$ is simply $2m$, the number of points in $\S$.  In three dimensions, the universal term in the entanglement entropy for spherical regions is identified with the well-known free energy $F$ on the sphere \cite{Casini:2011kv, Jafferis:2011zi, Klebanov:2011gs}.

In this paper we focus on four-dimensional (4D) CFTs in curved spacetime, where the universal term in the R\'enyi entropy \er{snuv} can be written as \cite{Fursaev:2012mp}
\be\la{sun}
S_n^{\univ} = \(\fr{f_a(n)}{2\pi} \cR_\S +\fr{f_b(n)}{2\pi} \cK_\S -\fr{f_c(n)}{2\pi} \cC_\S \) \ln\e \,.
\ee
Here $f_a$, $f_b$, and $f_c$ are coefficients that depend on $n$, and we have defined three conformal invariants
\bg\la{crc}
\cR_\S \eq \int_\S d^2y \sqrt{\g} R_\S \,,\qu
\cC_\S \eq \int_\S d^2y \sqrt{\g} \T C{ab}{ab} \,,\\\la{ck}
\cK_\S \eq \int_\S d^2y \sqrt{\g} \[\tr K^2 -\fr{1}{2} (\tr K)^2\] \,,
\eg
where $y$, $\g$, $R_\S$, and $K$ are the coordinates, induced metric, intrinsic Ricci scalar, and extrinsic curvature tensor of $\S$, and $\T C{ab}{ab}$ denotes the contraction of the Weyl tensor projected to directions orthogonal to $\S$.

Entanglement entropy can be studied by taking the $n\to1$ limit.  In this limit, the universal term is completely determined by the central charges of the CFT that appear in the Weyl anomaly \cite{Solodukhin:2008dh}:
\be
f_a(1) = a \,,\qu
f_b(1) = f_c(1) = c \,.
\ee

Away from $n=1$, the coefficients $f_a$, $f_b$, and $f_c$ are generally not determined from the central charges.  They depend on more physical data of the CFT.  It was noticed that $f_a$ can be extracted by considering a spherical entangling region, in which case it is determined by the thermal free energy of the CFT on a hyperboloid \cite{Hung:2011nu}.  The coefficient $f_c$ may be obtained by considering a small shape deformation and working to first order in the deformation.  This involves the stress tensor one-point function on the hyperboloid background, which is related to the thermal free energy.  In this way it was shown in \cite{Lewkowycz:2014jia} that $f_c$ is determined by $f_a$:
\be
f_c(n) = \fr{n}{n-1}\[ a -f_a(n) -(n-1)f_a'(n) \] \,.
\ee
It is also known that $f_b$ is in principle determined by working to second order in the shape deformation \cite{Lewkowycz:2014jia}.  Similar perturbative calculations were performed in other contexts in \cite{Rosenhaus:2014woa, Rosenhaus:2014zza, Allais:2014ata, Mezei:2014zla}.

The main goal of this paper is to determine $f_b$ by using gauge-gravity duality \cite{Maldacena:1997re, Gubser:1998bc, Witten:1998qj}.  Our basic strategy is to relate $f_b$ to the stress tensor one-point function in a deformed version of the hyperboloid background.  The latter is then determined by a straightforward holographic calculation.

It was conjectured in \cite{Lee:2014xwa} that
\be\la{conj}
f_b(n) = f_c(n)
\ee
is a universal property of all 4D CFTs for all $n$.  The evidence includes the surprising fact that it seems to hold in any free field theory involving an arbitrary number of scalars and fermions \cite{Lee:2014xwa}.  There have been recent attempts to prove or use this conjecture \cite{Lewkowycz:2014jia, Perlmutter:2015vma, Bueno:2015qya, Bueno:2015lza, Bianchi:2015liz}.  In particular, it was shown in \cite{Bueno:2015lza} to be equivalent to a conjectural relation between the universal contribution to the R\'enyi entropy from a small conical singularity on the entangling surface and the conformal dimension $h_n$ of the twist operator.  It was further shown in \cite{Bianchi:2015liz} that \er{conj} is equivalent to another conjecture relating $h_n$ to the two-point function of a displacement operator for twist operators.  However, we will prove here that this conjecture fails for holographic theories.  We will see this by calculating $f_b(n)$ either numerically for arbitrary $n$ or analytically by an expansion in $n-1$.

\se{R\'enyi entropy from the replica trick}

We use the replica trick to calculate the R\'enyi entropy
\be
S_n \eq \fr{1}{1-n} \ln \tr \r^n
\ee
of some region $A$ with the density matrix $\r$.  For an integer $n>1$, it may be obtained from
\be\la{renz}
S_n = \fr{\ln Z_n-n\ln Z_1}{1-n} \,,
\ee
where $Z_n$ is the partition function of the field theory on a suitable manifold known as the $n$-fold branched cover.

To study this concretely, we adopt a coordinate system similar to the Gaussian normal coordinates in a neighborhood of the entangling surface $\S$.  It is a codimension-2 surface, and on it we choose an arbitrary coordinate system $\{y^i, i=1, 2, \dots, d-2\}$.  From each point on $\S$ we may find a one-parameter family of geodesics orthogonal to $\S$.  Let us denote the parameter by $\t$ and employ the coordinates $(\r,\t,y^i)$ in a neighborhood of $\S$, where $\r$ is the radial distance to $\S$ along such a geodesic.  Choosing the parameter $\t$ judiciously \cite{Unruh:1989hy} so that its range is fixed as $2\pi$, we find that the metric in the neighborhood of $\S$ is
\be\la{met}
ds^2 = d\r^2 + G_{\t\t} d\t^2 + G_{ij} dy^i dy^j + 2G_{\t i} d\t dy^i \,,
\ee
where regularity at $\r=0$ requires the expansions
\ba\la{gtt}
G_{\t\t} &= \r^2\[1 + T \r^2 + \cO(\r^3)\] \,,\\\la{gij}
G_{ij} &= \g_{ij} + 2K_{aij} x^a + Q_{abij} x^a x^b + \cO(\r^3) \,,\\\la{gti}
G_{\t i} &= \r^2\[U_i + \cO(\r)\] \,.
\ea
Here $x^{1,2} \eq \r (\cos\t, \sin\t)$ are the coordinates orthogonal to $\S$, and Latin indices such as $a$ and $b$ denote these two directions, while $T$, $\g_{ij}$, $K_{aij}$, $Q_{abij}$, and $U_i$ are expansion coefficients that generally depend on $y^i$.  In particular, $\g_{ij}$ and $K_{aij}$ are the induced metric and extrinsic curvature tensor of $\S$.

Since the metric \er{met} is periodic under $\t \to \t+2\pi$, we may define a different manifold by extending the range of $\t$ from $2\pi$ to $2\pi n$ as long as $n$ is an integer.  This defines the $n$-fold branched cover.  It has a conical excess at $\r=0$ (i.e.\ the entangling surface $\S$), which we regulate by introducing a short distance cutoff at $\r=\e$.

It is useful to rewrite the conformal invariants appearing in \er{crc} and $\er{ck}$ as
\ba\la{ks}
& \tr K^2 -\fr{1}{2} (\tr K)^2 = K_{aij} K^{aij} -\fr{1}{2} K_{a} K^a \,,\\\la{cabab}
& \T C{ab}{ab} = \fr{R_\S}{3} -2T -\fr{2}{3} U_i U^i -\fr{1}{3} K_a K^a +\fr{1}{3} \Tud Q{a}{ai}{i} \,,
\ea
where $K_a \eq \Tu K{ai}i$ is the trace of the extrinsic curvature tensor.  We always use the induced metric $\g_{ij}$ to raise and lower Latin indices $i$, $j$ on $K$, $Q$, and $U$.

\se{Deformed hyperboloid}

For a spherical entangling region in the vacuum state of a CFT, the R\'enyi entropy can be determined by conformally mapping the problem to one of finding the free energy of the CFT on a unit hyperboloid with temperature $T=1/2\pi n$ \cite{Hung:2011nu}.  A spherical entangling surface has vanishing $\cK_\S$ and $\cC_\S$, so its R\'enyi entropy gives $f_a$ but not $f_b$ or $f_c$.  To obtain the latter two coefficients, we consider small shape deformations of $\S$ away from a perfect sphere.  It is most convenient to choose the undeformed entangling surface as a flat plane (i.e.\ a sphere with infinite radius) with
\be\la{gzero}
G_{\t\t}^{(0)} = \r^2 \,,\qu
G_{ij}^{(0)} = \d_{ij} \,,\qu
G_{\t i}^{(0)} =0 \,,
\ee
and treat terms in \er{gtt}--\er{gti} such as the extrinsic curvature $K$ as shape deformations.

We may perform an arbitrary Weyl transformation $g_{\m\n} = \W^2 G_{\m\n}$ on the metric \er{met} without affecting the R\'enyi entropy.  This is because the change of the partition function under a Weyl transformation is governed by the Weyl anomaly, which is an integral of local geometric invariants.  Such terms cancel between $\ln Z_n$ and $n \ln Z_1$ in the R\'enyi entropy \er{renz}, because locally the $n$-fold branched cover is identical to the original spacetime manifold on which the field theory is defined (away from the conical excess $\S$) \footnote{This is made precise by removing a small neighborhood of the conical excess $\S$, because the Weyl anomaly (including its boundary contributions) depends locally on the geometry.}.

Let us therefore consider the conformally equivalent metric $g_{\m\n} = G_{\m\n} / \r^2$:
\be\la{metp}
ds^2 = \fr{d\r^2 + G_{\t\t} d\t^2 + G_{ij} dy^i dy^j + 2G_{\t i} d\t dy^i}{\r^2} \,.
\ee
In the undeformed case \er{gzero}, the metric \er{metp} simplifies to
\be\la{hyp}
ds_{(0)}^2 = g_{\m\n}^{(0)} dx^\m dx^\n = d\t^2 + \fr{d\r^2 + \d_{ij} dy^i dy^j}{\r^2} \,,
\ee
which describes $\bH^{d-1} \times S^1$, a product of the $(d-1)$-dimensional hyperbolic space of unit radius and the $\t$ circle of size $2\pi n$.  For simplicity we call this product space the hyperboloid background and refer to it as $H_n^d$.

In the general case of \er{gtt}--\er{gti}, we view the metric \er{metp} as a deformed version of the hyperboloid background:
\be\la{dhyp}
g_{\m\n} = g_{\m\n}^{(0)} + \d g_{\m\n} \,.
\ee
We call this the deformed hyperboloid background and refer to it as $\wtd H_n^d$.

Our basic strategy for calculating the R\'enyi entropy is to perturbatively calculate the partition function on the deformed hyperboloid background using the fact that the change of the partition function is governed by the stress tensor one-point function:
\be\la{dzn}
\d \ln Z_n = \fr{1}{2} \int d^dx \sqrt{g} \<T^{\m\n}\> \d g_{\m\n} \,.
\ee

\se{$f_b$ from the stress tensor}

We now work in four dimensions and show that the coefficient $f_b$ is determined by the stress tensor one-point function in the deformed hyperboloid background to first order in the extrinsic curvature $K$.  Our basic idea is that \er{dzn} relates the second-order variation of the partition function to the first-order variation of the stress tensor one-point function.

Since our goal is to calculate $f_b$, we isolate it by turning on a small traceless extrinsic curvature tensor $K$.  It is clear from \er{cabab} that such a traceless $K$ does not contribute to $\T C{ab}{ab}$ or $\cC_\S$.  Neither does it contribute to $\cR_\S$, a topological invariant of the two-dimensional entangling surface.  Therefore, such a deformation allows us to easily extract $f_b$.  It is worth noting that we can always make $K$ traceless by performing a suitable Weyl transformation.  Therefore we realize a traceless $K$ perturbation by deforming the entangling surface away from a flat plane and applying an appropriate Weyl transformation to remove the trace of $K$.

In the hyperboloid background deformed by a traceless $K$, the stress tensor one-point function along the $y^i$ directions is
\be\la{tijp}
\<T^{ij}\>_{\wtd H_n^4} = \r^2 \[P_n \d^{ij} + \a_n \Tu Ka{ij} x^a + \cO(\r^2) \] \,,
\ee
where $P_n$ and $\a_n$ are $n$-dependent coefficients to be determined.  The first term $P_n \d^{ij}$ is the stress tensor one-point function in the perfect hyperboloid background, whereas the second term contributes to the universal term $\cK_\S$ in \er{sun} and determines $f_b$.

Inserting \er{tijp} into \er{dzn} with $\d g_{\m\n}$ given by a variation of the traceless extrinsic curvature
\be
\d g_{ij} = \fr{2\d K_{aij} x^a}{\r^2} \,,
\ee
we find
\ba\la{dznk}
\d \ln Z_n &= 2\pi n \a_n \int_\e \fr{d\r}{\r^3} \int_\S d^2y \sqrt{\g} \Tu Ka{ij} \d K_{bij} x^a x^b +\cO(\r^3) \nn\\
&= -\pi n \a_n \ln\e \int_\S d^2y \sqrt{\g} K^{aij} \d K_{aij} + \cd\,,
\ea
where the dots denote terms that are finite as $\e\to0$.  Here $\e$ plays the role of an infrared (IR) regulator on the infinite hyperboloid.  Integrating \er{dznk} in $K$, we obtain the $\cO(K^2)$ term in the logarithmically divergent part of the partition function
\be\la{znk}
\lt. \ln Z_n \rt|_{K^2} = -\fr{\pi n \a_n}{2} \ln\e \int_\S d^2y \sqrt{\g} K^{aij} K_{aij} \,.
\ee
Inserting this into \er{renz} and comparing it with \er{sun}, we arrive at
\be\la{fba}
f_b(n) = \pi^2n  \fr{\a_n -\a_1}{n-1} \,.
\ee
This result shows that $f_b$ is completely determined by the coefficient $\a_n$ appearing in the stress tensor one-point function \er{tijp} in the deformed hyperboloid background.

For completeness it is worth mentioning that the coefficient $f_c$ is determined by $P_n$ appearing in the stress tensor one-point function in the perfect hyperboloid background:
\be\la{fcp}
f_c(n) = -3\pi^2 n \fr{P_n -P_1}{n-1} \,.
\ee
This relation can be shown by using \er{dzn} with a shape deformation that affects $\cC_\S$ but not $\cK_\S$ \cite{Lewkowycz:2014jia}.  Considering for example the deformation given by $\d g_{ij} = Q_{abij} x^a x^b / \r^2$, we obtain
\be
\lt. \ln Z_n \rt|_{Q} = -\fr{\pi n P_n}{2} \ln\e \int_\S d^2y \sqrt{\g} \Tud Q{a}{ai}{i}\,.
\ee
Inserting this into \er{renz} and comparing it with \er{sun} with the help of \er{cabab}, we obtain \er{fcp}.

It is worth noting that the above results can be reproduced by similar calculations in the conical background \er{met}.  This involves reversing the Weyl transformation \er{metp} and finding the stress tensor one-point function in \er{met} from \er{tijp}.  There is an anomalous contribution which is analogous to the Schwarzian derivative in two-dimensional CFTs, but it depends locally on the geometry and cancels between $\ln Z_n$ and $n \ln Z_1$ in the R\'enyi entropy \er{renz}.

\se{Holographic calculation}

To obtain the coefficient $f_b$, we still need to calculate $\a_n$ in the stress tensor one-point function \er{tijp}.  Here we finish this last step using gauge-gravity duality.  Let us consider a holographic CFT dual to a gravitational theory in a bulk spacetime with one additional dimension.  The CFT lives on the asymptotic boundary of the bulk spacetime, and expectation values of local operators such as the stress tensor in the CFT are determined by the asymptotic behaviors of the corresponding fields such as the metric in the bulk.

The bulk metric that asymptotes to the deformed hyperboloid background \er{metp} is \footnote{We work in the units where the radius of curvature in the asymptotic bulk geometry is set to 1.}
\bm\la{hbh}
ds_{\bulk}^2 = \fr{dr^2}{f(r)} + f(r) d\t^2 \\
+ \fr{r^2}{\r^2} \Big\{ d\r^2 + \[\d_{ij} +2k(r) K_{aij}x^a\]dy^i dy^j \Big\} +\cd \,,
\em
where we have focused on deformations by a traceless extrinsic curvature tensor $K$, and the dots denote higher-order terms in $\r$.  This metric describes a deformed (Euclidean) hyperbolic black hole.  We choose the bulk coordinates using orthogonal geodesics originating from the black hole horizon (a codimension-2 surface), similar to the procedure described above \er{met}.  The metric \er{hbh} is uniquely fixed at this order in $\r$ by the bulk equations of motion up to diffeomorphisms.  In cases where the five-dimensional bulk is governed by Einstein gravity, the blackening factor $f(r)$ is
\be
f(r) = r^2 -1 -\fr{r_h^2 (r_h^2-1)}{r^2}
\ee
as determined by Einstein's equations in the metric \er{hbh} to leading order in $\r$.  Here $r_h$ is the location of the horizon and determined as a function of $n$ by the larger root of
\be\la{nrh}
n = \fr{2}{f'(r_h)} = \fr{r_h}{2r_h^2-1} \,.
\ee
To see this, we impose regularity at the horizon with the range of $\t$ being $2\pi n$.

Expanding Einstein's equations in the metric \er{hbh} to next order in $\r$, we find a second-order differential equation for the function $k(r)$:
\be\la{keq}
k''(r) + \[\fr{3}{r} +\fr{f'(r)}{f(r)}\] k'(r) - \fr{r^2 +f(r)}{r^2 f(r)^2} k(r) =0\,.
\ee
Generic solutions to this equation behave like $(r-r_h)^{\pm n/2}$ near the horizon.  Regularity of the extrinsic curvature deformation in \er{hbh} therefore demands $k(r) \sim (r-r_h)^{n/2}$ near $r=r_h$.  The solution to \er{keq} is uniquely determined by this IR boundary condition and the UV boundary condition $\lim_{r\to\infty} k(r)=1$.  Expanding the solution near the asymptotic boundary, we find
\be
k(r) = 1- \fr{1}{2r^2} +\fr{\b_n}{r^4} +\cO\(\fr{1}{r^6}\) \,,
\ee
where $\b_n$ is the coefficient of the normalizable mode and not fully determined by analysis near the asymptotic boundary.

The stress tensor one-point function in the CFT is determined by the asymptotic expansion of the bulk metric.  Using the results of \cite{deHaro:2000vlm}, we find \er{tijp} with
\be
P_n = \fr{\(r_h^2 -\fr{1}{2}\)^2}{16 \pi G_N} \,,\qu
\a_n = \fr{4\b_n -r_h^4 +r_h^2 +\fr{1}{4}}{8 \pi G_N} \,,
\ee
where $G_N$ denotes Newton's constant.  Inserting these values into \er{fba} and \er{fcp}, we obtain
\ba\la{fbr}
f_b(n) &= \fr{n (4\b_n -4\b_1 +r_h^2 -r_h^4)}{n-1} c \,,\\\la{fcr}
f_c(n) &= \fr{3n (r_h^2 -r_h^4)}{2(n-1)} c \,,
\ea
where we have used the relation $c = \pi/8G_N$ and that $r_h=1$ when $n=1$ according to \er{nrh}.

It remains to determine the coefficient $\b_n$.  We solve the differential equation \er{keq} numerically and plot the resulting $f_b(n)$ against $f_c(n)$ in Fig.~\re{figbc}.  They coincide at $n=1$ but not generally.  It is worth noting that their difference is quite small for a large range of values of $n$, raising the question of whether the numerical proof of $f_b(n) = f_c(n)$ in \cite{Lee:2014xwa} for free field theories was established with sufficient accuracy \footnote{We thank Eric Perlmutter and Aitor Lewkowycz for discussions on this point.}.

\begin{figure}[h]
\centering
\includegraphics[width=\columnwidth]{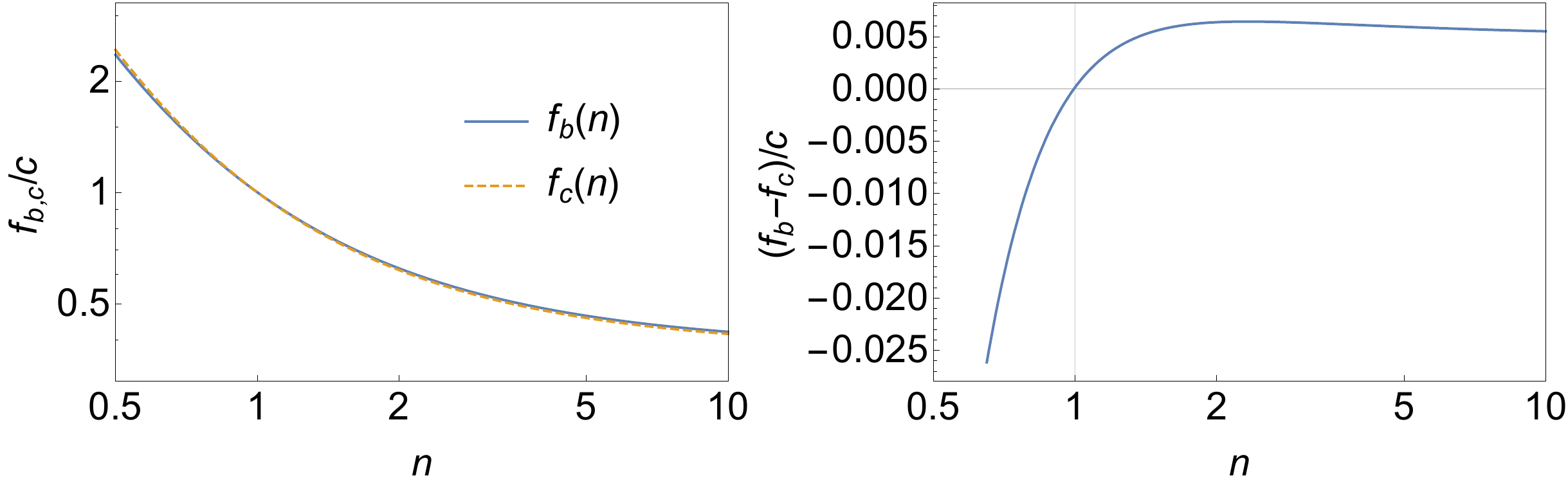}
\caption{Plots of $f_b(n)$ against $f_c(n)$ in units of the central charge $c$ in holographic CFTs.  In the left logarithmic plot we show both of them for the range $0.5 \leq n \leq 10$.  We show their difference more clearly in the right plot.}
\la{figbc}
\end{figure}

Alternatively, we can solve \er{keq} perturbatively in $n-1$.  To do this we define
\be
h(r) \eq k(r) \exp\[\int_r^\infty \fr{dr}{f(r)}\] \,,
\ee
and the differential equation $\er{keq}$ becomes
\be
h''(r) +\[\fr{3}{r} +\fr{2+f'(r)}{f(r)}\] h'(r) +\fr{3r-1}{r^2f(r)} h(r)=0 \,.
\ee
The advantage of working with $h(r)$ is that the regularity condition at the horizon simply requires $h(r_h)$ to be finite.  Expanding in $n-1$, we find
\be
h(r) = \fr{r+1}{r} +(n-1) h_1(r) +(n-1)^2 h_2(r) +\cd \,,
\ee
where
\ba
&h_1(r) = \fr{r+1}{r} \ln\(\fr{r+1}{r}\) -\fr{6r^2 +3r -1}{6r^3} \,,\\
&\bmd
h_2(r) = \fr{r+1}{2r} \ln^2 \(\fr{r+1}{r}\) \\
\!\!\! -\fr{6r^2 +3r -1}{6r^3} \ln\(\fr{r+1}{r}\) +\fr{216 r^3 -85 r +27}{432 r^5} \,.
\emd
\ea
From the asymptotic behaviors of these functions we obtain
\be
\b_n = -\fr{1}{8} +\fr{n-1}{12} -\fr{67(n-1)^2}{432} +\cO(n-1)^3 \,.
\ee
Inserting this into \er{fbr} we arrive at
\be
f_b(n) = \[1 -\fr{11}{12} (n-1) +\cO(n-1)^2\] c \,,
\ee
which agrees with
\be
f_c(n) = \[1 -\fr{17}{18} (n-1) +\cO(n-1)^2\] c
\ee
when $n=1$ but not for general $n$.

Similar perturbative techniques can be used in the small $n$ limit:
\be
f_b(n) = \fr{1+\cO(n)}{16 n^3}c \,,\qu
f_c(n) = \fr{3+\cO(n)}{32 n^3}c \,,
\ee
or in the large $n$ limit, leading to
\be
f_b(n) \ap 0.3800c +\cO(n^{-1}) \,,\qu
f_c(n) = \fr{3}{8}c +\cO(n^{-1}) \,.
\ee

\se{Discussion}

The universal coefficient $f_b$ governs the variation of the R\'enyi entropy under traceless extrinsic curvature deformations in 4D CFTs.  We have seen that it is entirely determined by the stress tensor one-point function in the deformed hyperboloid background, which we have calculated holographically.  Surprisingly, our results disprove the $f_b(n) = f_c(n)$ conjecture.  It is worth exploring why this relation seems to hold for free field theories but fails holographically.

The coefficient $f_b$ is not only related to the stress tensor one-point function, but is also connected to the universal contribution to the R\'enyi entropy from a conical entangling surface and the two-point function of a displacement operator for twist operators.  A more general conjecture, proposed in two equivalent ways in \cite{Bueno:2015lza, Bianchi:2015liz} for an arbitrary CFT in any dimensions, relates the universal conical contribution and the displacement operator two-point function to the conformal dimension of the twist operator.  This conjecture is equivalent to $f_b(n) = f_c(n)$ in four dimensions and therefore is also disproved by our holographic results.  However, it is worth studying this conjecture in other dimensions, either using an analog of the techniques developed here or applying the area-law prescription for holographic R\'enyi entropy recently proposed in \cite{Dong:2016fnf}.

It is worth exploring why the violation of the $f_b(n) = f_c(n)$ conjecture in holographic CFTs appears small for a large range of values of $n$.  It opens up the possibility that the conjecture holds approximately and provides a simple method of calculating $f_b$ from $f_c$ with reasonable accuracy.  Finally, our results form a step towards studying the shape dependence of entanglement and R\'enyi entropies in many other contexts and dimensions.

\se{Acknowledgments}

\begin{acknowledgments}
I would like to thank Pablo Bueno, Aitor Lewkowycz, Juan Maldacena, Robert Myers, Eric Perlmutter, and William Witczak-Krempa for useful discussions, and the Stanford Institute for Theoretical Physics where this work was started.  This work was supported in part by the National Science Foundation under Grant No. PHY-1316699, by the Department of Energy under Grant No. DE-SC0009988, and by a Zurich Financial Services Membership at the Institute for Advanced Study.
\end{acknowledgments}

\bibliography{bibliography}

\end{document}